\documentclass[twocolumn,prl,aps,superscriptaddress,showpacs]{revtex4}

\usepackage{graphicx}
\usepackage{dcolumn}
\usepackage{amsmath}
\newlength{\figwidth}
\newcommand{\bra}[1]{\left<#1\right\vert}
\newcommand{\ket}[1]{\left\vert#1\right>}

\begin{document}


\title{Probing high-energy electronic excitations using inelastic neutron scattering}
\author{Young-June Kim}
\email{yjkim@physics.utoronto.ca} \affiliation{Department of
Physics, University of Toronto, Toronto, Ontario M5S~1A7, Canada}

\author{A. P. Sorini}
\affiliation{Stanford Institute for Materials and Energy Science,
SLAC National Accelerator Laboratory, Menlo Park, California 94025,
USA}

\author{C.~Stock}
\affiliation{NIST Center for Neutron Research, 100 Bureau Drive,
Gaithersburg, Maryland 20899, and Indiana University Cyclotron
Facility, 2401 Milo B. Sampson Lane, Bloomington, Indiana 47404,
USA.}

\author{T.~G.~Perring}
\affiliation{ISIS Facility, Rutherford Appleton Laboratory, STFC,
Chilton, Didcot, Oxon OX11 0QX, and Department of Physics and
Astronomy, University College London, Gower Street, London WC1E 6BT,
United Kingdom}

\author{J. van den Brink}
\affiliation{Institute for Theoretical Solid State Physics, IFW
Dresden, 01171 Dresden, Germany}

\author{T. P. Devereaux}
\affiliation{Stanford Institute for Materials and Energy Science,
SLAC National Accelerator Laboratory, Menlo Park, California 94025,
USA} \affiliation{Geballe Laboratory for Advanced Materials,
Stanford University, Stanford, California 94305, USA}

\date{\today}

\begin{abstract}
High-energy, local multiplet excitations of the d-electrons are
revealed in our inelastic neutron scattering measurements on the
prototype magnetic insulator NiO. These become allowed by the
presence of both non-zero crystal field and spin-orbit coupling. The
observed excitations are consistent with optical, x-ray, and EELS
measurements of d-d excitations. This experiment serves as a proof
of principle that high-energy neutron spectroscopy is a reliable and
useful technique for probing electronic excitations in systems with
significant crystal field and spin-orbit interactions.
\end{abstract}



\maketitle

Electrons in strongly correlated systems are characterized by the
dual nature of their spatial wave functions; sometimes they are
considered localized, but other times delocalized. This dichotomy
makes it difficult for traditional band theory to describe physical
properties of these scientifically interesting and technologically
relevant materials. Often a model based on local interactions, such
as a Hubbard Hamiltonian, is used to capture the main physics of
these systems, but distilling the numerous interactions present in a
condensed matter system into a simple model requires information
from experiments. For this reason, various types of electron
spectroscopy tools have played significant roles in elucidating
various interactions such as crystal field (CF) and spin-orbit (SO)
coupling. These include optical, x-ray absorption, photoemission,
and electron energy loss spectroscopy (EELS) as well as inelastic
x-ray scattering (IXS) in recent years. Here we show that inelastic
neutron scattering (INS) can be added to this arsenal of
spectroscopic tools for probing {\em electronic structure} of
correlated electron systems.

Neutron spectroscopy has been used to study crystal field splittings
in rare-earth insulators\cite{brockhouse62} and metals
\cite{rainford68,turberfield70,birgeneau71,fulde85}. With the
availability of copious amounts of epithermal neutrons from
spallation neutron sources, even inter-multiplet transitions in
f-electron systems have been observed with INS
\cite{shapiro75,taylor88}. Despite earlier theoretical predictions
of using INS in studying interband transitions in semiconductors
\cite{cooke82}, no experimental studies of electronic excitations
above 1 eV have been reported to date. Only recently, it was shown
that quantitative information above 1 eV can be obtained from direct
geometry spectrometers \cite{Stock10}. We note that novel IXS
methods have been introduced utilizing either quadrupole transition
matrix element \cite{larson07,haverkort07} or resonant enhancement
in the soft \cite{ghiringhelli05,chiuzbaian05} or hard x-ray range
\cite{huotari08}. However, such charge-sensitive probes are not
ideal for investigating doped Mott insulators, such as cuprate
superconductors, because charge carrier contributions dominate the
excitation spectrum in the energy range of interest. On the other
hand, charge-neutral neutrons do not couple to charge degrees of
freedom, and could be very valuable for measuring d-d excitations in
such metallic compounds.

In this Letter, as a first step towards this goal, we report our
observation of localized d-d excitations in NiO using INS. NiO is a
prototypical Mott insulator with a large charge-transfer gap of
about 4 eV. This material has been studied extensively using various
spectroscopic techniques, and its electronic structure is quite well
known. In particular, many novel spectroscopic tools have been
recently applied to study NiO d-d excitations as a test case, since
the NiO d-d excitations are well characterized \cite{newman59}. For
example, spin-polarized EELS \cite{fromme95}, soft x-ray resonant
inelastic x-ray scattering (RIXS)
\cite{ghiringhelli05,chiuzbaian05}, hard x-ray RIXS
\cite{huotari08}, non-resonant IXS \cite{larson07}, and
angle-resolved EELS \cite{muller08} have all been used to detect d-d
excitations in NiO. The benefit of studying NiO is that there is no
ambiguity with regard to the identification of d-d excitations, thus
allowing us to focus on the mechanism of the spectroscopy.

In our INS investigation of NiO, d-d excitations at 1.0 eV and 1.6
eV have been observed, which is the highest energy solid-state
excitation observed with INS to date \cite{taylor88}. These
transitions occur via magnetic dipole and higher order scattering
operators, respectively, giving strong momentum dependence to their
spectral weights. We find that either CF splitting or SO coupling is
necessary for the observed transitions, suggesting that one could
determine the strengths of these interactions using momentum
dependent INS spectrum. We also studied single magnon density of
states, which peak up near the zone boundary energy. In contrast to
recent soft x-ray RIXS studies, we do not find atomic spin flip
scattering at the expected energy range \cite{ghiringhelli09}. Our
observation thus opens up an exciting new field of electron
spectroscopy using INS at quite an opportune time; recently
commissioned Spallation Neutron Source will undoubtedly provide an
excellent venue for such studies.


Total 250 g powder sample of NiO (Alfa Aesar 99 \%) was studied on
the MAPS and MARI direct geometry time of flight spectrometers
located at ISIS (Rutherford Appleton Labs, U.K.).  On MAPS, an
incident energy of 3 eV was selected by spinning the nimonic chopper
at 100 Hz and the ``A" Fermi chopper at 600 Hz. On MARI, an incident
energy of 500 meV was used with a 50 Hz nimonic and a 600 Hz ``A"
chopper. All measurements were done at room temperature. Note that
the Neel temperature of NiO is well above room temperature.

\begin{figure}
\begin{center}
\includegraphics[angle=0,width=3.25in]{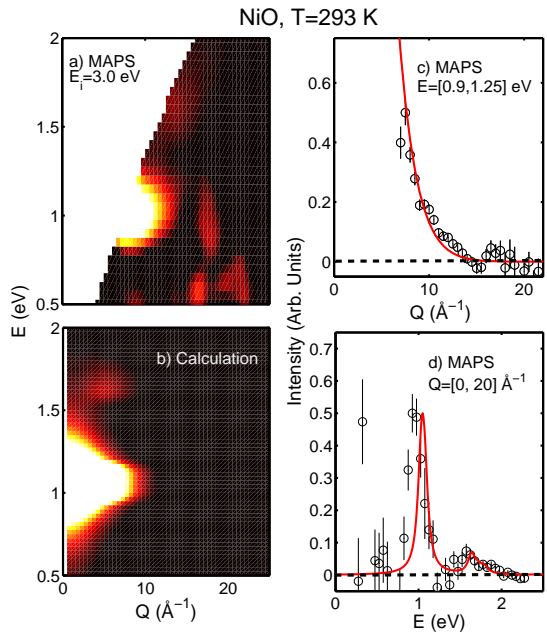}
\end{center}
\vspace*{-0.8cm} \caption{(Color online) Panel a) is the magnetic
intensity map of energy-momentum space obtained on MAPS using
incident neutron energy of 3 eV at room temperature. The background
intensity due to phonons has been subtracted as discussed in the
text. b) illustrates the theoretical calculation discussed in the
text. Panels c) and d) illustrate integrated cuts in momentum and
energy as noted. The open circles are experimental data and the
solid (red) lines are theoretical calculations. As described in the
text, the overall amplitude is the only adjustable parameter in the
calculation.} \label{fig1}
\end{figure}

In Fig. 1(a), the intensity map of the excitation spectrum of NiO
obtained with $E_i$=3 eV is plotted. The horizontal axis is momentum
transfer (${\bf Q}$) and the vertical axis is energy transfer. We
have subtracted the high-Q background from the raw data. It is
believed that the high Q data mostly consist of contributions due to
multiple phonons, and the high-Q data were fitted to a form of
$A+BQ^2$ and subtracted from the low-Q data as described in
Ref.~\cite{stock-nmo}. One can clearly identify a strong spectral
feature around the excitation energy of 1 eV, which is more or less
dispersionless. The spectral weight on the other hand shows quite
strong Q dependence and the intensity falls to the background level
around Q=15 \AA$^{-1}$, which is clearly seen in the integrated
intensity plot shown in Fig.~1(c). The intensity at higher momentum
is due to the background fluctuation. In addition to the 1 eV
feature, one can also notice a faint excitation feature around 1.6
eV. This can be more clearly seen in Fig. 1(d), in which we plot the
intensity integrated up to Q= 20 \AA$^{-1}$ as a function of energy.
The well-defined peak centered around 1 eV and additional intensity
around 1.6 eV are identified as d-d excitations.

To check if there are hydrogens in the sample, we measured potential
hydrogen recoil line \cite{Stock10}. In Fig.~2, we plot the energy
and momentum region where hydrogen recoil line is expected (white
solid line). As shown in this figure, there is no signature of any
hydrogen scattering, eliminating the possibility that the observed
excitations come from hydrogen.

\begin{figure}
\begin{center}
\includegraphics[angle=0,width=3in]{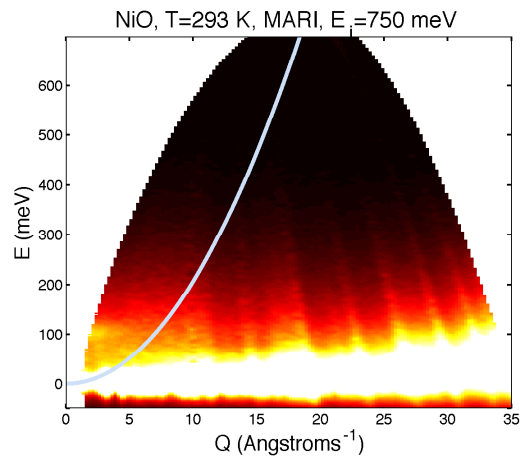}
\end{center}
\vspace*{-0.5cm} \caption{(Color online) Intensity map of expanded
energy and momentum space. The solid white line would correspond to
hydrogen recoil.} \label{fig2}
\end{figure}


In order to understand the scattering mechanism and the observed
momentum dependence, we have carried out a cluster calculation. We
consider the electronic system to be described by a local model
\cite{degrootcf} in which the relevant degrees of freedom are the
d-electrons of the Ni$^{2+}$ ion in an octahedral ligand field. Our
choice of model parameters\cite{chiuzbaian05,ghiringhelli05} are
consistent with previous theory and experiments on d-d excitations
of NiO, and were not adjusted to fit our data \footnote{The model is
parameterized\cite{griffith61} by the $t_{2g}-e_g$ splitting
$10Dq=1.05$~eV, the SO coupling \cite{cowanbook} $\xi= 0.051$~eV,
and the Racah parameters \cite{cowanbook,griffith61} $B=0.11$~eV and
$C=0.42$~eV.}. The model includes explicit electron-electron
interactions and is solved using the exact-diagonalization method to
obtain all the eigenstates and eigenenergies.

The eigenstates and energies are used via Fermi's golden rule
to obtain the double differential scattering
cross-section for unpolarized neutrons as
\begin{eqnarray}\label{eq:ddcs_full}
\frac{d\sigma}{d\Omega d\omega}&=& \frac{k_f}{k_i}
\left(\frac{\gamma_n\mu_N}{c}\right)^2 \sum_f {\left| {\bf v}
\right|}^2 \delta(E_0+\omega-E_f)\;,
\end{eqnarray}
where\cite{balcar89}
\begin{equation}\label{eq:v}
{\bf v}= \frac{1}{Q^2}{\bf Q} \times \bra{\Psi_f} \sum_i e^{i{\bf
Q}\cdot{\bf r_i}}\left({\bf p_i}+i{\bf s_i}\times{\bf Q}\right)
\ket{\Psi_0}\;,
\end{equation}
${\bf Q}={\bf k}_i-{\bf k}_f$ is the momentum transferred to the
electrons, $\omega$ is the energy lost by the neutron, and
$\ket{\Psi_n}$ and $E_n$ are the states and energies of the
electronic system.

The matrix elements in Eq.~(\ref{eq:v}) have a significant
dependence on the CF and SO coupling, which can be illustrated by
considering the simple limit where $Q\to 0$ (i.e., the
magnetic-dipole approximation). In this approximation the allowed
excitations are connected to the ground state by the operator ${\bf
L}+2{\bf S}$, where ${\bf L}$ is the total orbital angular momentum
and ${\bf S}$ is the total spin. For an isolated non-relativistic
atom neither ${\bf L}$ nor ${\bf S}$ can cause inelastic
transitions. However, in the presence of a CF it is well known that
the diagonal matrix elements of ${\bf L}$ are ``quenched". The
weight that is removed from these diagonal matrix elements
effectively appears in the off-diagonal matrix elements; the
presence of a CF allows ${\bf L}$ to contribute to the inelastic
scattering signal. Similarly, the presence of SO coupling allows
${\bf S}$ to contribute to the inelastic signal. The magnetic dipole
approximation illustrates the importance of CF and SO.

However, due to neutron kinematics, the momentum transfers at
energy-loss appropriate for d-d excitation (on the order of eV) can
be large compared to the inverse radial size of the electronic
orbitals. Thus the magnetic-dipole approximation is not always valid
and the full expression, Eq.~(\ref{eq:v}), must be retained. We will
see that the dipolar terms give rise to a non-dispersive
intra-multiplet peak at $\omega\approx 10Dq$ which dominates the
spectrum at low momentum transfer. For higher momentum transfers,
non-dipole excitations appear at higher energies (e.g., 1.6~eV) and
eventually dominate the relative scattering intensity. The origin of
these peaks can be understood roughly from the Tanabe-Sugano diagram
\cite{tanabe54} for a $d^8$ system, which describes the change in
multiplet energies as a function of CF strength; transitions from
the $A_1$ octahedral symmetry ground state to the higher energy
$T_2$ and $T_1$ character states give rise to the spectral peaks at
$\sim 1.0$~eV, and $\sim 1.6$~eV, respectively. The locations of
these peaks are slightly shifted and split by SO coupling. The
relative intensity of each peak changes with the magnitude of $Q$ as
determined by the matrix-elements of Eq.~(\ref{eq:v}).

\begin{figure}
\includegraphics[angle=0,width=3.25in]{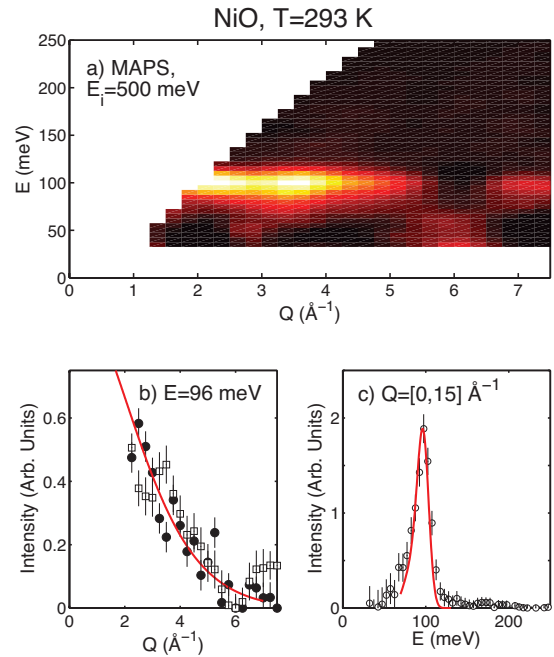}
\vspace*{-0.5cm}
\caption{(Color online) Panel a) illustrates a false contour plot
taken on MAPS with the E$_{i}$=500 meV.  b) the intensity of the
peak at 96 meV as a function of momentum transfer.  The solid and
open points are from MARI (E$_{i}$=750 meV) and MAPS (E$_{i}$=500
meV) respectively.  The solid red line is the free ion Ni$^{2+}$
form factor.  c) The momentum integrated scattering and solid line
is the calculated density of states from Ref.~\cite{Hutchings72b}.}
\label{fig3}
\end{figure}

In Fig.~1(b) we show the theoretical INS intensity map. The energy
integral from 0.9 to 1.25 eV over this map, and the momentum
integral for $Q<20$~\AA$^{-1}$, are also shown in Fig.~1(c)-(d) as
solid lines. The agreement between theory and experiment is quite
good, given that the only adjustable parameter is the overall
amplitude. In particular, the momentum dependence shown in Fig.~1(c)
suggests that the 1 eV transition occurs via magnetic dipole
operator, ${\bf L}+2{\bf S}$. The 1.6 eV transition spectral weight
is only non-zero in the intermediate Q range, and vanishes both at
small and large Q. Since SO coupling in this compound is small, the
multiplet splitting due to the SO coupling only shows up as
broadening of the 1.6 eV peak.

In addition to the high energy spectrum, we also measured the low
energy spin excitation spectrum with $E_i$=500 meV, as shown in
Fig.~3. A clear magnon mode is observed around 100 meV, which is the
zone boundary spin wave energy. The assignment of this feature is
unambiguous in INS. Earlier neutron scattering studies of spin wave
dispersion reported that the zone boundary magnon energy is 117 meV
\cite{Hutchings71,Hutchings72b}. In Fig. 3(b), we plot the intensity
as a function of Q. The solid line is the magnetic form factor from
the spin angular momentum contribution as tabulated in
Ref.~\cite{Brown95}. In Fig. 3(c), the low-Q intensity (integrated
up to 5 \AA$^{-1}$.) is plotted as a function of energy transfer.
Since we are using a powder sample, this integration is a quick
method to obtain the magnon density of states. We compare this
result with the calculated magnon density of states from Hutchings
and Samuelsen \cite{Hutchings72b}, which agrees very well.

These lower-energy results shed light on understanding excitations
observed with different spectroscopic techniques in the energy range
of 100-200 meV. As shown in Fig.~3, no significant scattering
intensity is observed above the one magnon band up to 250~meV. This
should be compared with earlier photon spectroscopy investigation.
In optical spectroscopy studies \cite{dietz71}, a two-magnon mode at
193~meV was observed with Raman scattering, while an infrared
absorption mode due to two-magnon plus phonon was observed at
250~meV. In a recent Ni L$_3$-edge RIXS study \cite{ghiringhelli09},
the excitations at 95 meV and 190 meV were associated with
transitions between atomic levels split by an exchange field
\cite{degroot98}. We note that ``two-magnon" excitations in neutron
scattering usually refers to longitudinal spin fluctuations, which
usually forms a broad continuum and has very small intensity
\cite{huberman05}. The apparent lack of excitations around 190~meV
in our INS data clearly illustrates the difference between the
``two-magnon" excitations detected with photon spectroscopy and
neutron spectroscopy. Further theoretical calculation is necessary
to address this issue quantitatively \cite{Chen11}.

In summary, we observed d-d excitations of 1 eV and 1.6 eV in NiO
using inelastic neutron scattering. Such local multiplet transitions
are allowed in the magnetic dipole and higher order channel due to
the presence of non-zero crystal field and spin-orbit coupling. The
excitation energy matches well with previous experiments, and the
momentum dependence of the spectral weight could be well described
by the cluster model calculation. Our observation illustrates that
inelastic neutron scattering at high energies is a reliable and
useful technique for probing electronic excitations in materials
with significant crystal field and spin-orbit coupling.

We would like to thank K. Plumb for valuable discussions. Y. K. was
supported by Natural Sciences and Engineering Research Council of
Canada. A.P.S. and T.P.D. are supported by the U.S. Department of
Energy under Contracts No. DE-AC02-76SF00515 and No.
DE-FG02-08ER46540 (CMSN).

\end{document}